# Third Harmonic Generation in Transparent Longitudinal Epsilon-Near-Zero Multilayers

Wallace Jaffray[1], Sven Stengel[1], Domenico de Ceglia[2], Neset Akozbek[3], Filippo Capolino[4], Marcello Ferrera[1], Michael Scalora[2] and Maria Antonietta Vincenti[2,*]

[1]Institute of Photonics and Quantum Sciences, Heriot-Watt University, Edinburgh, EH14 4AS, UK;
[2]Department of Information Engineering, Università degli Studi di Brescia, Brescia, Italy;
[3]U.S. Army Space and Missile Defense Command, Redstone Arsenal, 35898 – AL (USA);
[4]Department of Electrical Engineering and Computer Science, University of California, Irvine, California 92697, USA;
*maria.vincenti@unibs.it

**Abstract:** Epsilon-near-zero (ENZ) materials can dramatical-ly enhance local optical fields, enabling nonlinear interactions at relatively low intensities. Yet, near their plasma frequency, conventional isotropic ENZ media remain highly absorptive, limiting nonlinear operations that require good transparency. Longitudinal epsilon-near-zero (LENZ) metamaterials, characterized by a vanishing permit-tivity along the optical axis provide an exceptional plat-form for field enhancement while mitigating absorption losses and impedance mismatch. We experimentally show that an αSi/ITO multilayer engineered for a LENZ resonance in the near-infrared enables broadband, high pump transmission while still harnessing ENZ-enhanced nonlinearity to generate a strong third-harmonic signal. This demonstrates that efficient nonlinear processes can be driven without the high-loss conditions typical of isotropic ENZ media and regardless of intrinsic absorption at the harmonic frequency. The resulting third harmonic efficiency is comparable to isotropic ENZ films but without the absorption-induced heating constraints of ENZ operation. The high pump transmission enables transparent LENZ (TLENZ) stacks to be integrated into optical cavities, where resonant field buildup could amplify the nonlinear response without compromising thermal management. These results establish TLENZ multilayers as a robust, versatile platform for transparent, field-enhanced nonlinear nanophotonics, combining strong light–matter interaction with low-loss operation.

## Introduction

Over the past decades, materials with vanishing permittivity, or epsilon-near-zero (ENZ) media, have captured interest across the photonics community for their ability to dramatically reshape light propagation. As the real part of the dielectric permittivity approaches zero, optical wavelengths effectively stretch toward infinity inside the medium, producing a nearly uniform internal phase, dramatically enhanced light–matter coupling, and exotic wave phenomena such as supercoupling and highly directive emission [1-7]. Beyond linear control of electro-magnetic waves, both natural and engineered ENZ materials have emerged as potent amplifiers of nonlinear optical interactions. In these systems, exceptional field confinement drives efficient harmonic generation and enables extreme nonlinear dynamics such as soliton formation, optical bistability and ultrafast switching [8-26]. However, applications in isotropic ENZ media, like transparent conducting oxides (TCOs), are limited by their relevant absorption near the plasma frequency [27, 28]. This is further aggravated as the conditions that produces maximum field enhancement coincides with a peak in the material absorption, thus preventing transparent nonlinear regimes. This limitation has driven interest in anisotropic ENZ materials, where a single element of the permittivity tensor vanishes.

 By alternating an ENZ material with another isotropic dielectric, one realizes a longitudinal ENZ (LENZ) medium in which the longitudinal component of the effective permittivity goes to zero while the transverse component remains finite [29]. LENZ designs are predicted to deliver much larger intensity enhancement and dramatically broader angular acceptance than their isotropic counterparts. Those advantages have been exploited to boost second-harmonic generation efficiencies by more than an order of magnitude [30]. More broadly, extreme anisotropy promotes enhanced nonlinear interactions, especially in hybrid metal–TCOs architectures [31].

A similar strategy appears in metal–dielectric multilayers, where alternating thin metallic and dielectric films exposes the metal's intrinsic nonlinearity and avoids the heavy absorption of a bulk metal [32-35]. In these systems, careful engineering of the layer thicknesses allows nonlinear processes such as harmonic generation to be activated while maintaining significant optical transmission [36-39]. This analogy underlines how judicious layering offers a general pathway to combine strong nonlinearity with transparency.

Despite these theoretical promises, experimental confirmation of LENZ-enhanced nonlinear optics has remained elusive. Harmonic generation has been reported in anisotropic ENZ systems such as phonon-polariton AlN [40], but those studies did not directly probe harmonic generation in the LENZ regime or examine the interplay between nonlinearity and transparency.

In the present work, we close this gap by demonstrating THG in a TLENZ multilayer. We design and fabricate an indium tin oxide (ITO) and amorphous silicon (aSi) multilayer stack that supports a longitudinal ENZ resonance in the near infrared while simultaneously showing broadband transparency. Thanks to the unique combination of TLENZ properties we can produce a strong nonlinear response without entering the high-loss pump regime that limits the use of isotropic ENZ media. The observed third harmonic signal is in excellent agreement with our theoretical predictions both qualitatively and quantitatively. Moreover, efficient THG persists despite substantial absorption at the harmonic frequency, owing to phase-locking between the pump and the third harmonic [41-43]. These results experimentally validate the ENZ route to strong nonlinear optics under transparent operation and establish TLENZ multilayers as a robust platform that combines intense light–matter interactions with low-loss operation.

**Effective permittivity response analysis**

We designed a ten-periods stack composed of alternating layers of αSi and ITO [Fig.1(a)] to demonstrate the ability to combine a transparent linear response with efficient nonlinear processes. Individually, αSi and ITO are isotropic materials (see SI, Note 1), when layered periodically with subwavelength thicknesses, the composite behaves as an effectively anisotropic system. Using Maxwell–Garnett effective medium theory, the transverse (in-plane, x and y) and longitudinal (out-of-plane, z) components of the permittivity tensor are set by the constituent permittivities and the ITO volume filling ratio (FR). The transverse component is the arithmetic average, $[\varepsilon_t = FR\varepsilon_{\text{ITO}} + (1-FR)\varepsilon_{\alpha\text{Si}}]$, while the longitudinal component follows a harmonic-type average $[\varepsilon_l = \frac{\varepsilon_{\text{ITO}}\varepsilon_{\alpha\text{Si}}}{FR\varepsilon_{\alpha\text{Si}} + (1-FR)\varepsilon_{\text{ITO}}}]$. In the near infrared, aSi is a high-index dielectric with positive permittivity, while ITO crosses from positive to negative-permittivity near its plasma frequency at 1248 nm. Consequently, small changes in the relative thicknesses within the multilayer can drive the effective longitudinal permittivity through zero, reversing the sign of the optical response and producing a large contrast between longitudinal and transverse components. Fig.2 shows the real and imaginary parts of the effective permittivities as a function of both wavelength and FR for a aSi/ITO stack with period p = 50 nm. The white dots in Fig.2(a) mark the wavelength-FR combination that gives $\varepsilon_l = 0$. Increasing the ITO filling ratio above 20% barely shifts the zero-crossing wavelength, but the degree of anisotropy grows as the filling ratio decreases. Around the zero-crossing wavelength (~1260$nm$) the real part of the transverse effective permittivity $\varepsilon_t$ increases as the ITO fraction falls, while the imaginary parts $Im(\varepsilon_t)$ stays low and spectrally featureless, whereas $Im(\varepsilon_l)$ spikes just below the ITO plasma frequency for low FR [see Fig.2(b)]. Overall, the control of $\epsilon_t$ with the FR around 1260$nm$ seen Fig.2(c) offers an extra degree of freedom to control transparency and impedance matching, regardless of the materials above and below the multilayer.

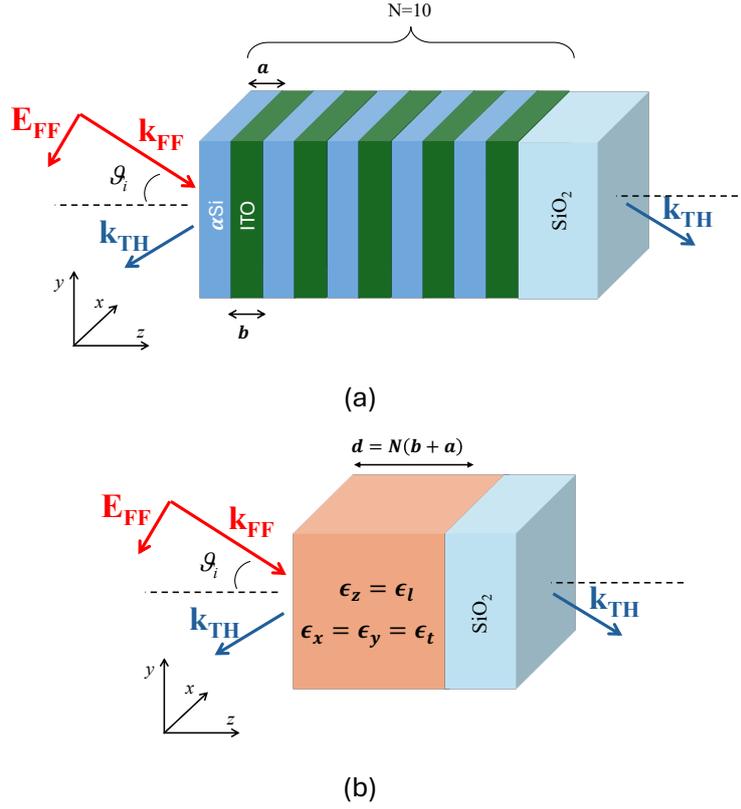

**Fig. 1:** (a) Sketch of the multilayer and pump excitation scheme. The multilayer is composed by N=10 periods of aSi (thickness *a*) and ITO (thickness *b*) so that each period is equal to p = 50 nm for a total thickness of the multilayer d = N(*b*+*a*) = 500 nm. (b) Effective medium representation of the multilayer in (a).

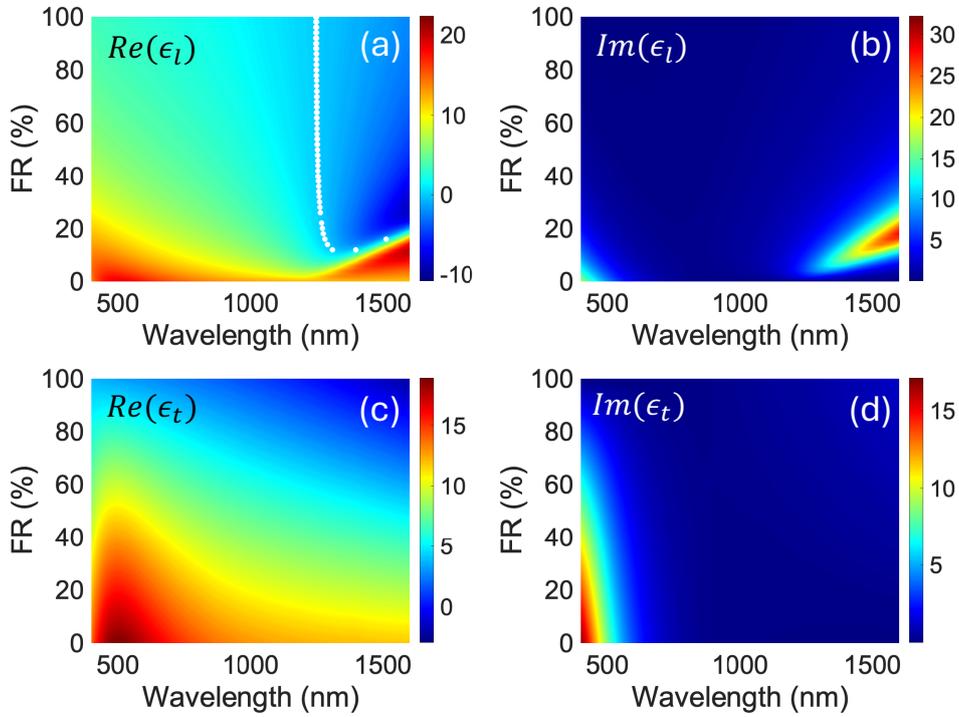

**Fig. 2:** Effective permittivities of a αSi/ITO multilayer with period p = 50 nm as a function of wavelength and ITO filling ratio (FR). (a) Real part of the longitudinal permittivity $Re[\varepsilon_l]$, the white dots mark the wavelength/FR pairs where $Re[\varepsilon_l] = 0$. The zero-crossing wavelength remains largely unchanged above a 20% ITO FR, while decreasing the FR enhances the anisotropy. (b) Imaginary part of the longitudinal permittivity $Im[\varepsilon_l]$; a pronounced

peak emerges just above the ITO plasma frequency at low FRs. (c) Real part of the transverse permittivity $Re[\varepsilon_t]$, which increases near the zero-crossing wavelength (~1260 nm) as the ITO FR decreases. (d) Imaginary part of the transverse permittivity $Im[\varepsilon_t]$, which remains low and spectrally featureless across the parameter range.

**Combining transparency with efficient nonlinear response**

Strong anisotropy, combined with a near-zero longitudinal permittivity, can markedly boost the field intensity enhancement and broaden the angular acceptance compared to isotropic ENZ media, as shown for LENZ [29]. While such field enhancement is essential for driving nonlinear processes, our goal is also to retain high transparency at the pump wavelength, a condition unattainable in isotropic ENZ [27] or for a generic combination of materials that form a LENZ condition [31]. To meet both requirements, we fabricated a TLENZ made of ten periods of aSi and ITO with a FR=50% (for fabrication details see SI, Note 2). Maxwell-Garnett theory gives effective permittivities at $\lambda = 1263 nm$ of $\varepsilon_l = 0.009 + i2.396$ and $\varepsilon_t = 8.992 + i0.204$. According to Ref [29], anisotropy is responsible for field intensity enhancement (FIE). Therefore, under the condition $|\varepsilon_l| \ll \sin\theta_i$, where $\theta_i$ is the angle of incidence, one has $\text{FIE} \approx 4|\epsilon_t|^2/|\epsilon_l|^2 \cos^2\theta_i$; despite such ideal condition does not reflect our operating conditions, this formula readily shows in a simple way the beneficial effect of anisotropy. In other words, it is not important to have a small $|\epsilon_l|$ but rather a large $|\epsilon_t|^2/|\epsilon_l|^2$ as we have in the designed TLENZ. We then measured the stack's linear transmittance by utilizing a stable and calibrated white-light source placed at almost normal incidence to the multilayer device to collect the normalised power transmission using a visible and near-infrared spectrometer [Fig. 3, red markers]. We optimized multilayer transmission with a transfer-matrix model, starting from 25nm nominal thickness per layer (FR=50%) and assuming the ITO and aSi dielectric permittivities identical to those retrieved with ellipsometric measurements on the single layers (see SI, Note 1). The model was iteratively fitted to the measured transmittance by allowing each layer thickness to vary within ±5 nm. Searching this restricted parameter space, we adjusted individual layer thicknesses to minimize the difference between simulated and experimental transmission, yielding an optimized stack that best reproduces the observed optical response [Fig. 3, blue line]. We stress how the measured transmittance exceeds 50% across a broad spectral range, indicating that ITO and aSi layers indeed form coupled cavities that resonate and permit significant transmission despite the overall thickness d of the stack and the material losses at play [29-32].

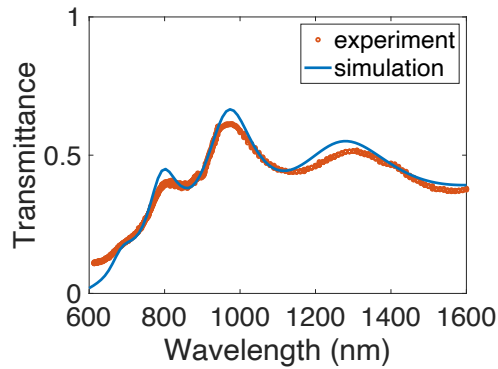

**Fig. 3:** Transmission spectra (Transmittance) of the 10 period aSi/ITO multilayer stack as measured (red markers) and simulated (blue line).

We then moved forward to measure the TH signal by pumping the multilayer at 1250nm. We first measured the angular response with a pump intensity I = 100 GW cm⁻². and then we measured THG for increasing pump intensities. Specifically, the samples were excited using near-infrared femtosecond pulses generated by an optical parametric amplifier (OPA) with wavelength tuneable in the 1150-1450 nm range. Pump power was adjusted to produce on-sample intensities in the range of 15–150 GW cm⁻². The beam was focused to a 0.35 mm diameter spot and operated at a 10 Hz repetition rate with 85 fs pulse duration. The generated

TH signal was collected in transmission and analysed spectrally using a calibrated detection system (see SI, Note 3).

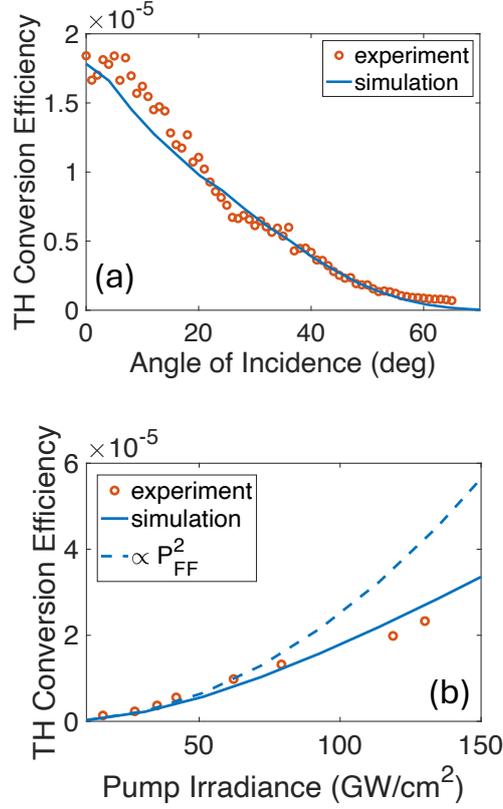

**Fig. 4:** (a) Measured (red markers) and simulated (blue line) TH conversion efficiency of the ten period aSi/ITO multilayer stack as a function of the angle of incidence for a pump wavelength at 1250 nm and 100 GW/cm² pump intensity. (b) Measured (red markers) and simulated (solid blue line) TH conversion efficiency for increasing pump intensity and fixed pump wavelength at 1250 nm for normal incidence. Dashed line shows the predicted power scaling law of THG without including the terms in Eq.(1).

We solved the THG problem at fixed pump wavelength of 1250 nm using the finite element method (COMSOL Multiphysics) by implementing the nonlinear polarization terms for the pump at ω and its TH at 3ω. We assume an isotropic third-order nonlinear response for both ITO and aSi, using the nonlinear susceptibilities and dispersion reported in [44]. For the crystal symmetry considered, only three independent tensor components of $\chi^{(3)}$ contribute, so the nonlinear polarization reduces to its scalar form [45]. The nonlinear polarization induced by the pump field $E_\omega$ at ω and 3ω are written as

$$P^{NL}_{\omega,i} = \epsilon_0 \chi^{(3)} E_{\omega,j} \left( 3|E_{\omega,j}|^2 + 2|E_{\omega,k}|^2 + 2|E_{\omega,l}|^2 \right) \quad (1)$$

$$P^{NL}_{3\omega,i} = \epsilon_0 \chi^{(3)} E_{\omega,j} \left( E^2_{\omega,j} + E^2_{\omega,k} + E^2_{\omega,l} \right) \quad (2)$$

These expressions include only pump–pump and pump–pump–pump mixing terms that drive self-phase modulation of the fundamental frequency and for THG. Other frequency-mixing contributions are neglected because they do not contribute significantly to the processes. The model reproduces the measured third harmonic conversion efficiency both qualitatively and quantitatively when considering both variability with respect to angle of incidence [Fig.4(a)] and pump intensity [Fig.4(b)]. Conversion efficiency here is defined as the ratio of third-harmonic output power to the fundamental input power. We note how the simulation correctly predicts the sub-quadratic scaling of THG efficiency with pump intensity only when Eq.(1) and (2) are solved simultaneously with the finite element method. Omitting Eq. (1) or solving the two equations in segregated steps produces the near-quadratic trend expected from a simplified theory [Fig.4(b), dashed, blue line]. The model developed with the finite element method through Eq.(1) and (2) assumes a continuous wave pump tuned at 1250nm and is not sufficient to reproduce the TH response in a pulsed regime, nor to include nonlocal effects that might be appreciated in a time domain dynamics. For these reasons we use a time-domain theory that fully couples matter dynamics to the macroscopic Maxwell's equations. Examples of the method can be found in references [32] and [46]. Concretely, we use a hydrodynamic-Maxwell model designed to capture both linear and nonlinear optics of conducting oxides and amorphous silicon. The model simultaneously accounts for surface, magnetic, and bulk nonlinear sources from free and bound charges; retains linear and nonlinear dispersion; includes nonlocal corrections from pressure and viscosity; allows the free-electron effective mass in ITO to change under strong absorption; and can incorporate pump depletion when local fields become sufficiently large. The standard macroscopic treatment of conducting oxides retains only the free-electron (Drude) term. For the time-domain analysis we model ITO's dielectric function as the sum of a Drude background and a single ultraviolet Lorentzian resonance, so the local permittivity acquires an added bound-electron contribution as follows:

$$\epsilon_{ITO}(\omega) = 1 - \frac{\omega_p^2}{\omega^2 + i\gamma_f \omega} - \frac{\omega_{p,b}^2}{\omega^2 - \omega_{o,b}^2 + i\gamma_b \omega} \qquad (3)$$

In Eq.(3), $\omega_p^2 = 4\pi n_{o,f} e^2/m_f^*$ is the plasma frequency; $n_{o,f}$ the free electron density; $e$ the electronic charge; $\gamma_f$ the free electron damping coefficient; and $m_f^*$ the temperature-dependent effective free electron mass; $\omega_{p,b}^2$ is the bound electron plasma frequency, defined similarly to the free electron counterpart; $\omega_{o,b}^2$ is the resonance frequency; and $\gamma_b$ the bound electron damping coefficient. Including the Lorentzian response injects a crucial, dynamic degree of freedom into the nonlinear response. This approach in fact produces an intrinsic, dispersive resonant nonlinearity in the ITO that is absent from purely Drude descriptions and can affect harmonic generation. Practically, two nonlinear channels can compete. One is the resonant, dispersion-mediated response tied to the Lorentzian bound resonance, the other is the nonlinearity that arises when free carriers heat and the effective mass (and hence the Drude response) evolves with pump intensity. Their relative importance depends on spectral detuning, material dispersion, and incident pump intensity. This combined Drude-Lorentz description therefore modifies both linear phase and the frequency dependence of the nonlinear susceptibility, with direct consequences for harmonic-generation performance. Our model can capture altered phase matching conditions, bandwidth, and conversion efficiency, and can either enhance or suppress harmonics produced by free-carrier mechanisms. Ignoring the Lorentzian terms risks misattributing spectral features or the scaling of harmonic yield with pump power. Including it yields a more faithful, predictive dielectric function across the experimental spectral range and helps identify which physical pathway (resonant bound-electron nonlinearity or free-carrier heating) dominates in each design, enabling more informed choices to maximize nonlinear efficiency while controlling dispersion and loss. Unlike ITO, which we treat with a Drude background plus a UV Lorentzian, amorphous silicon is modelled as a dielectric composed of three Lorentzian resonances, each carrying its own intrinsic, dispersive nonlinear response. Building on these assumptions, we formulate coupled, time-domain equations of motion that yield two distinct polarization channels: one from free carriers and one from bound electrons, written separately for each material [44]. We neither impose a preformed dielectric constant, nor introduce shortcuts (for example, pseudo-singular denominators because of the ENZ crossing point) to rationalize large conversion efficiencies. Instead, fitting a retrieved local dielectric function only fixes physical parameters (damping rates, effective masses, and carrier densities) that will populate the dynamical equations; the linear dielectric constant can be reconstructed *a posteriori*, if desired. For both materials in our stack the polarization equations for free ($P_f$) and bound ($P_b$) electrons include the lowest-order bulk, surface, and magnetic nonlinear sources, which can account for even harmonic generation in centrosymmetric materials like ITO and $\alpha$Si. These terms are adapted from prior work on metal and semiconductor nonlinear optics [12, 44] and adapted here to capture temperature- and time-dependent carrier dynamics. Plasma and resonance frequencies in both ITO and $\alpha$Si are allowed to evolve with temperature and intensity (we assume the total number of free carriers remains constant, although their local density can vary in space and time), so absorption-driven changes in effective mass and density feedback self-consistently into linear dispersion and nonlinear coupling. This unified formulation exposes competing nonlinear channels, i.e., resonant Lorentzian contributions from bound electrons in $\alpha$Si and ITO, free-carrier dynamics in ITO, nonlocal and surface effects, and lets their relative importance emerge from the dynamics rather than from prescriptive assumptions. The time-domain dynamics of free and bound charges can, therefore, be written as:

$$\ddot{\boldsymbol{P}}_f + \gamma_f \dot{\boldsymbol{P}}_f = \frac{n_{o,f} e^2}{m_f^*(T_e)}\boldsymbol{E} - \frac{e}{m_f^*(T_e)}\boldsymbol{E}(\nabla \bullet \boldsymbol{P}_f) + \frac{e}{m_f^*(T_e)}\dot{\boldsymbol{P}}_f \times \boldsymbol{H} + \frac{3E_F}{5m_f^*(T_e)}\left(\nabla(\nabla \bullet \boldsymbol{P}_f) + \nabla^2 \boldsymbol{P}_f\right) - \frac{1}{n_{o,f} e}\left[(\nabla \bullet \dot{\boldsymbol{P}}_f)\dot{\boldsymbol{P}}_f + (\dot{\boldsymbol{P}}_f \bullet \nabla)\dot{\boldsymbol{P}}_f\right] \tag{4}$$

$$\ddot{\boldsymbol{P}}_{bj} + \gamma_{bj}\dot{\boldsymbol{P}}_{bj} + \omega_{o,bj}^2 \boldsymbol{P}_{bj,NL} = \frac{n_{o,bj} e^2}{m_{bj}^*}\boldsymbol{E} - \frac{e}{m_{bj}^*}(\boldsymbol{P}_{bj} \bullet \nabla)\boldsymbol{E} + \frac{e}{m_{bj}^* c}\dot{\boldsymbol{P}}_{bj} \times \boldsymbol{H} \tag{5}$$

In the free-carrier equation [Eq.(4)] $T_e$ denotes the electron temperature; in the bound-carrier equation [Eq.(5)] $n_{o,bj}$ and $m_{bj}^*$ denote the bound-electron mass and density, both assumed to be constant; $\boldsymbol{P}_{bj}$ represents the jth bound-electron polarization for a given material, and $\boldsymbol{P}_{bj,NL} = \alpha \boldsymbol{P}_{bj}\boldsymbol{P}_{bj} - \beta(\boldsymbol{P}_{bj} \bullet \boldsymbol{P}_{bj})\boldsymbol{P}_{bj} + \cdots$ is the nonlinear polarization. The tensors $\alpha$ and $\beta$ encode crystal symmetry (for centrosymmetric media like ITO and amorphous Si one has $\alpha = 0$). The third order nonlinear coefficient is written for an isotropic medium but can be generalized to other symmetries. This coupled dynamical formulation intentionally omits any presupposed dielectric constant or refractive index. In our model linear and nonlinear dispersions, surface harmonic sources, nonlocal pressure and viscous terms, magnetic contributions, and convective terms all emerge from the charge dynamics rather than from mere assumptions. In particular, the free-electron polarization in ITO mirrors that of noble metals except that the relevant free-carrier parameters (for example the plasma frequency) depend on $T_e$. The same framework can readily accommodate time-varying free-carrier densities produced by interband excitation in metals or semiconductors.

Term by term, Eq. (4) groups Coulombic contributions, the magnetic Lorentz force, nonlocal pressure and viscosity terms, and finally convective nonlinearities. Eq. (5) is dominated by the bound-electron resonance (bound plasma frequency) and includes surface and magnetic Lorentz contributions. Because both equations retain lowest-order bulk, surface, and magnetic nonlinear sources, the model lets competing nonlinear channels (resonant bound responses, free-carrier heating and mass changes, nonlocal spatial dispersion, and surface harmonics at oblique incidence) reveal themselves self-consistently from the dynamics rather than from imposed singularities or fitted denominators. Moreover, we model hot carriers with a dynamical effective mass picture that captures the essential physics of a full two temperature model while remaining compact and transparent. For the modest electron temperatures relevant here (a few thousand °K), we approximate the ITO effective mass with the temperature set by absorption and the current density emerging from the free carrier equation of motion:

$$m_f^*(T_e) \approx m_o^* + \alpha K_B T_e = m_o^* + \alpha K_B \Lambda \iint \boldsymbol{J} \cdot \boldsymbol{E} d\boldsymbol{r}^3 dt \tag{6}$$

For convenience, this relation is simplified by defining $\boldsymbol{J} = \sigma_0 \boldsymbol{E}$, where $\sigma_0$ is a constant to be determined, which allows one to write in simplified form

$$\frac{n_{o,f} e^2}{m_f^*(T_e)} \boldsymbol{E} \approx \frac{n_{o,f} e^2}{m_0^*} \boldsymbol{E} - \widetilde{\Lambda}(\boldsymbol{E} \cdot \boldsymbol{E})\boldsymbol{E} + \widetilde{\Lambda}^2(\boldsymbol{E} \cdot \boldsymbol{E})^2 \boldsymbol{E} + \cdots \tag{7}$$

Using Eq.(6), the free electron plasma frequency in Eq.(4) is simplified to an intensity dependent law with a single fitted constant that incorporates absorption strength, interaction volume, and pulse duration. Using this temperature dependent effective mass, the dominant Coulomb term in the free carrier equation acquires an explicit intensity dependent shift that appears as a redshift of the plasma frequency and generates cubic, quintic and higher order nonlinearities that can become important with increasing peak power densities. Geometric factors and the pulse temporal overlap are folded into scalar coefficients, so that our simulations account for an intensity driven free carrier dynamic. The bound electron equation of motion remains unchanged. We then solve the free and bound electron equations together with the temperature-dependent effective mass and sum the resulting polarizations which is then inserted into Maxwell's equations. This approach retains the dominant physical route by which absorption alters dispersion and nonlinearity in the ITO schematically as follows: absorption heats carriers, effective mass changes, the plasma frequency redshifts, and intensity dependent behaviour follows. Therefore, we preserve the mechanisms required to capture sub-quadratic harmonic scaling, phase locking, and pump depletion effects when fields become large, while avoiding the complexity of a full two temperature problem solver. Higher order corrections and a full two temperature treatment can be reintroduced when experimental conditions demand it.

Fig. 5 compares the experimentally measured transmitted TH spectra [Fig. 5(a)] with the spectra predicted by our model [Fig. 5(b)], as the central wavelength of the pumping wave is tuned. The solid black trace in Fig. 5(b) shows the transmitted THG conversion efficiency from the simulation, its amplitude matches the measured values to within experimental uncertainty. This close qualitative and quantitative agreement indicates that the time-domain, hydrodynamic treatment, with coupled free- and bound-electron dynamics, temperature-dependent effective mass, and retained surface, magnetic, and nonlocal contributions, faithfully captures the dominant physical phenomenology that determine harmonic spectral response and harmonic efficiencies. Reproducing both spectral shape and efficiency confirms the model's ability to predict intensity-dependent scaling and phase-locking effects and supports its use for designing TLENZ multilayers that optimize conversion while preserving pump transparency. Finally, Fig. 6 shows the TH conversion efficiency as a function of pump wavelength for two peak incident intensities. At low intensity (10 GW/cm²), a THG peak of order $10^{-8}$ (right red axis) appears near ITO's ENZ wavelength, which remains stable around 1260 nm.

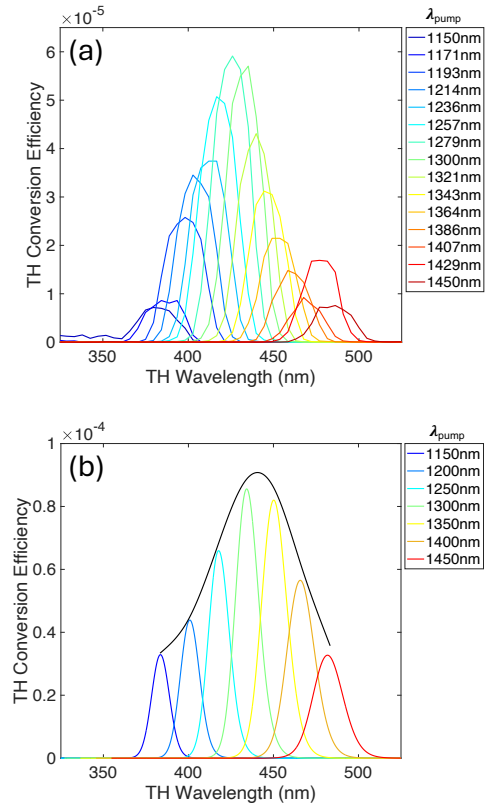

**Fig. 5:** (a) Measured and (b) simulated TH conversion efficiency of the ten period aSi/ITO multilayer stack as a function of the pump wavelength for 300 GW/cm² pump intensity. Simulations have been performed in the time domain using the hydrodynamic-Maxwell model.

By contrast, raising the peak power to 300 GW/cm² drives a near-100 nm redshift of the ITO plasma edge and boosts the predicted conversion efficiency by about four orders of magnitude (left blue axis). The plot illustrates two coupled effects. At weak excitation the ENZ resonance delivers modest field enhancement with negligible spectral shift, so conversion stays small. At higher intensities, absorption heats carriers, the effective mass and plasma frequency move, and the resonant condition migrates spectrally. This dynamical redshift is accompanied by strong nonlinear gain seen in the high-power curve. The result highlights how intensity-dependent dispersion in TLENZ media can convert modest ENZ enhancement into orders-of-magnitude gains in harmonic yield, while also highlighting the trade-offs introduced by pump-driven material shifts.

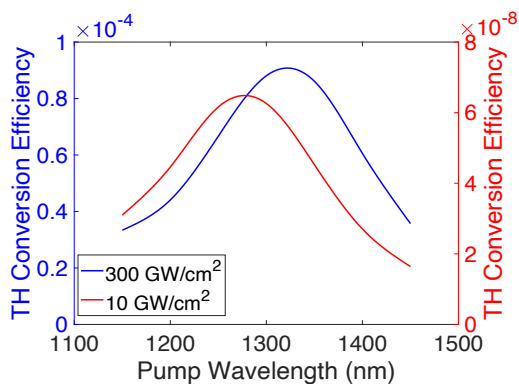

**Fig. 6:** Simulated TH conversion efficiency response as a function of wavelength for pump intensities of 10 GW/cm² (red line – right axis) and 300 GW/cm² (blue line – left axis).

## Conclusions

We have demonstrated efficient third-harmonic generation in a ten-period ITO/αSi multilayer, with both the finite element method and a fully coupled time-domain hydrodynamic model accurately reproducing the measured amplitudes and spectra. The stack maintains high pump transmission (>50%) while sustaining strong longitudinal field enhancement, showing that TLENZ architectures can reconcile low-loss operation with high nonlinear efficiency. Operating in a transparent pump regime is particularly advantageous: reduced absorption leads to improved thermal stability and permits substantially longer interaction lengths, offering an additional route to efficiency scaling beyond field enhancement alone. Our modelling derives the optical response from coupled microscopic equations for free and bound charges. This approach captures linear and nonlinear dispersion, nonlocal pressure and viscous terms, surface and magnetic sources, and an intensity-dependent effective mass describing hot-carrier dynamics, enabling predictive agreement with experiment without imposing *ad hoc* dielectric functions. The model identifies resonant bound-electron response, carrier heating, and nonlocality as the key factors governing the nonlinear yield and the observed sub-quadratic scaling. The TLENZ platform supports strong longitudinal fields with broad angular acceptance while keeping the pump in a low-loss window. The resulting third-harmonic efficiency is comparable to isotropic ENZ films but without the absorption and heating constraints of ENZ operation. The high pump transmission further enables the use of TLENZ stacks inside optical cavities, where resonant buildup could amplify the nonlinear response without compromising thermal management. Overall, TLENZ multilayers provide a robust, designable foundation for transparent nonlinear nanophotonics, combining tunable dispersion engineering, with low-loss cavity architectures.